\documentstyle[12pt]{article}
\advance \topmargin by -\headheight
\advance \topmargin by -\headsep

\evensidemargin \oddsidemargin
\begin{document}

\topmargin -.6in

\renewcommand{\thesection}{\Roman{section}.}
\renewcommand{\thesubsection}{\Roman{section}.\Roman{subsection}.}
\renewcommand{\thesubsubsection}{\Roman{section}.\Roman{subsection}.\Roman{subsubsection}.}
\newcommand{\sect}[1]{\setcounter{equation}{0}\section{#1}}
\renewcommand{\theequation}{\arabic{section}.\arabic{equation}}
\relax

%%      MACROS.TEX
%
%               macros formatting and equations
\def\rf#1{(\ref{eq:#1})}
\def\lab#1{\label{eq:#1}}
\def\nonu{\nonumber}
\def\br{\begin{eqnarray}}
\def\er{\end{eqnarray}}
\def\be{\begin{equation}}
\def\ee{\end{equation}}
\def\eq{\!\!\!\! &=& \!\!\!\! }
\def\foot#1{\footnotemark\footnotetext{#1}}
\def\lb{\lbrack}
\def\rb{\rbrack}
\def\llangle{\left\langle}
\def\rrangle{\right\rangle}
\def\blangle{\Bigl\langle}
\def\brangle{\Bigr\rangle}
\def\llb{\left\lbrack}
\def\rrb{\right\rbrack}
\def\lcurl{\left\{}
\def\rcurl{\right\}}
\def\({\left(}
\def\){\right)}
\newcommand{\nit}{\noindent}
\newcommand{\ct}[1]{\cite{#1}}
\newcommand{\bi}[1]{\bibitem{#1}}
\def\lskip{\vskip\baselineskip\vskip-\parskip\noindent}
\relax
\def\mskp{\par\vskip 0.3cm \par\noindent}
\def\sskp{\par\vskip 0.15cm \par\noindent}
%                     common physics symbols
\def\tr{\mathop{\rm tr}}
\def\Tr{\mathop{\rm Tr}}
\def\v{\vert}
\def\bv{\bigm\vert}
\def\Bgv{\;\Bigg\vert}
\def\bgv{\bigg\vert}
\newcommand\partder[2]{{{\partial {#1}}\over{\partial {#2}}}}
\newcommand\funcder[2]{{{\delta {#1}}\over{\delta {#2}}}}
\newcommand\Bil[2]{\Bigl\langle {#1} \Bigg\vert {#2} \Bigr\rangle}  %% <.|.>
\newcommand\bil[2]{\left\langle {#1} \bigg\vert {#2} \right\rangle} %% <.|.>
\newcommand\me[2]{\left\langle {#1}\right|\left. {#2} \right\rangle} %% <.|.>
\newcommand\sbr[2]{\left\lbrack\,{#1}\, ,\,{#2}\,\right\rbrack}
\newcommand\pbr[2]{\{\,{#1}\, ,\,{#2}\,\}}
\newcommand\pbrs[3]{\{\,{#1}\, ,\,{#2}\,\}_{#3}}
\newcommand\pbbr[2]{\lcurl\,{#1}\, ,\,{#2}\,\rcurl}
%
%                    math symbols
\def\a{\alpha}
\def\b{\beta}
\def\d{\delta}
\def\D{\Delta}
\def\eps{\epsilon}
\def\vareps{\varepsilon}
\def\g{\gamma}
\def\G{\Gamma}
\def\grad{\nabla}
\def\h{{1\over 2}}
\def\l{\lambda}
\def\L{\Lambda}
\def\m{\mu}
\def\n{\nu}
\def\o{\over}
\def\om{\omega}
\def\O{\Omega}
\def\p{\phi}
\def\P{\Phi}
\def\vp{\varphi}
\def\pa{\partial}
\def\pr{\prime}
\def\ra{\rightarrow}
\def\s{\sigma}
\def\S{\Sigma}
\def\t{\tau}
\def\th{\theta}
\def\Th{\Theta}
\def\ti{\tilde}
\def\wti{\widetilde}
\def\bp{{\bar \p}}
\newcommand\sumi[1]{\sum_{#1}^{\infty}}   %% summation till infinity
\newcommand\fourmat[4]{\left(\begin{array}{cc}  %%   2x2 matrix
{#1} & {#2} \\ {#3} & {#4} \end{array} \right)}
\newcommand\twocol[2]{\left(\begin{array}{c}  %%   2 column
{#1} \\ {#2} \end{array} \right)}

%
%%%                    macros for Lie algebras
\def\lie{{\cal G}}
\def\dlie{{\cal G}^{\ast}}
\def\f#1#2#3 {f^{#1#2}_{#3}}
\def\winf{{\sf w_\infty}}
\def\win1{{\sf w_{1+\infty}}}
\def\hwinf{{\sf {\hat w}_{\infty}}}
\def\Winf{{\sf W_\infty}}
\def\Win1{{\sf W_{1+\infty}}}
\def\hWinf{{\sf {\hat W}_{\infty}}}
\def\cB{{\cal B}}
\def\cH{{\cal H}}
\def\cL{{\cal L}}
\def\cA{{\cal A}}
\def\cK{{\cal K}}
\def\cM{{\cal M}}
\def\cR{{\cal R}}
\def\cP{{\cal P}}
\def\cg{{\cal G}}
%
%       fake blackboard bold macros for reals, complex, etc.
\def\rlx{\relax\leavevmode}
\def\inbar{\vrule height1.5ex width.4pt depth0pt}
\def\IZ{\rlx\hbox{\sf Z\kern-.4em Z}}
\def\IR{\rlx\hbox{\rm I\kern-.18em R}}
\def\IC{\rlx\hbox{\,$\inbar\kern-.3em{\rm C}$}}
\def\one{\hbox{{1}\kern-.25em\hbox{l}}}
%%
%
%       This defines the journal citations
%
\newcommand\PRL[3]{{\sl Phys. Rev. Lett.} {\bf#1} (#2) #3}
\newcommand\NPB[3]{{\sl Nucl. Phys.} {\bf B#1} (#2) #3}
\newcommand\NPBFS[4]{{\sl Nucl. Phys.} {\bf B#2} [FS#1] (#3) #4}
\newcommand\CMP[3]{{\sl Commun. Math. Phys.} {\bf #1} (#2) #3}
\newcommand\PRD[3]{{\sl Phys. Rev.} {\bf D#1} (#2) #3}
\newcommand\PLA[3]{{\sl Phys. Lett.} {\bf #1A} (#2) #3}
\newcommand\PLB[3]{{\sl Phys. Lett.} {\bf #1B} (#2) #3}
\newcommand\JMP[3]{{\sl J. Math. Phys.} {\bf #1} (#2) #3}
\newcommand\PTP[3]{{\sl Prog. Theor. Phys.} {\bf #1} (#2) #3}
\newcommand\SPTP[3]{{\sl Suppl. Prog. Theor. Phys.} {\bf #1} (#2) #3}
\newcommand\AoP[3]{{\sl Ann. of Phys.} {\bf #1} (#2) #3}
\newcommand\PNAS[3]{{\sl Proc. Natl. Acad. Sci. USA} {\bf #1} (#2) #3}
\newcommand\RMP[3]{{\sl Rev. Mod. Phys.} {\bf #1} (#2) #3}
\newcommand\PR[3]{{\sl Phys. Reports} {\bf #1} (#2) #3}
\newcommand\AoM[3]{{\sl Ann. of Math.} {\bf #1} (#2) #3}
\newcommand\UMN[3]{{\sl Usp. Mat. Nauk} {\bf #1} (#2) #3}
\newcommand\FAP[3]{{\sl Funkt. Anal. Prilozheniya} {\bf #1} (#2) #3}
\newcommand\FAaIA[3]{{\sl Functional Analysis and Its Application} {\bf #1}
(#2) #3}
\newcommand\BAMS[3]{{\sl Bull. Am. Math. Soc.} {\bf #1} (#2) #3}
\newcommand\TAMS[3]{{\sl Trans. Am. Math. Soc.} {\bf #1} (#2) #3}
\newcommand\InvM[3]{{\sl Invent. Math.} {\bf #1} (#2) #3}
\newcommand\LMP[3]{{\sl Letters in Math. Phys.} {\bf #1} (#2) #3}
\newcommand\IJMPA[3]{{\sl Int. J. Mod. Phys.} {\bf A#1} (#2) #3}
\newcommand\AdM[3]{{\sl Advances in Math.} {\bf #1} (#2) #3}
\newcommand\RMaP[3]{{\sl Reports on Math. Phys.} {\bf #1} (#2) #3}
\newcommand\IJM[3]{{\sl Ill. J. Math.} {\bf #1} (#2) #3}
\newcommand\APP[3]{{\sl Acta Phys. Polon.} {\bf #1} (#2) #3}
\newcommand\TMP[3]{{\sl Theor. Mat. Phys.} {\bf #1} (#2) #3}
\newcommand\JPA[3]{{\sl J. Physics} {\bf A#1} (#2) #3}
\newcommand\JSM[3]{{\sl J. Soviet Math.} {\bf #1} (#2) #3}
\newcommand\MPLA[3]{{\sl Mod. Phys. Lett.} {\bf A#1} (#2) #3}
\newcommand\JETP[3]{{\sl Sov. Phys. JETP} {\bf #1} (#2) #3}
\newcommand\JETPL[3]{{\sl  Sov. Phys. JETP Lett.} {\bf #1} (#2) #3}
\newcommand\PHSA[3]{{\sl Physica} {\bf A#1} (#2) #3}
\newcommand\PHSD[3]{{\sl Physica} {\bf D#1} (#2) #3}
\def\cKP{{\sf cKP}~}
\def\bfs{{\bf s}}
\def\bfts{{\bf {\tilde s}}}
\def\kere{\mbox{\rm Ker (ad $E$)}}
\def\ime{\mbox{\rm Im (ad $E$)}}
\def\qs{Q_{\bfs}}
\def\cgh{{\widehat {\cal G}}}

\begin{titlepage}
% \vspace{-1cm}
% \noindent
January, 1997 \hfill{IFT-P/006/97}\\
%\phantom{bla}
%\hfill{UICHEP-TH/95-9}\\
\phantom{bla}
\hfill{solv-int/9701012}
 \\
\vskip .3in

\begin{center}

{\large\bf Some comments on the bi(tri)-Hamiltonian structure}
\end{center}
\begin{center}
{\large\bf of Generalized AKNS and DNLS hierarchies{\footnotemark
\footnotetext{Work supported in part by CNPq}}}
\end{center}
\normalsize
\vskip .4in

\begin{center}
H.S. Blas Achic{\footnotemark
\footnotetext{Supported by FAPESP}}, L.A. Ferreira,
J.F. Gomes and A.H. Zimerman 

\par \vskip .1in \noindent
Instituto de F\'{\i}sica Te\'{o}rica-UNESP\\
Rua Pamplona 145\\
01405-900 S\~{a}o Paulo, Brazil
\par \vskip .3in

\end{center}

\begin{center}
{\large {\bf ABSTRACT}}\\
\end{center}
\par \vskip .3in \noindent

We give the correct prescriptions for the terms involving 
$\pa_x^{-1}\, \d (x-y)$,  in the Hamiltonian structures of the AKNS and DNLS 
systems, in order for the Jacobi identities to hold. We also establish that the 
$sl(2)$ AKNS and DNLS systems are tri-Hamiltonians and construct two compatible 
Hamiltonian structures for the $sl(3)$ AKNS system. 
We also give a derivation of the recursion operator for the $sl(n+1)$ DNLS
system.

\end{titlepage}

The Hamiltonian structures of the Generalized AKNS and DNLS systems were 
studied in refs. \ct{FK83,agz,liu} and \ct{fordy} respectively. In ref.
\ct{agz} it was obtained, in a closed form, the recurrence relation for 
consecutive time evolutions for the Generalized AKNS hierarchy, as well as the
first and second Hamiltonian structures. Ref. \ct{liu} discussed the
compatibility of those two structures. However, some special attention must be
paid to the second bracket, since it contains terms involving 
$\pa_x^{-1}\, \d (x-y)$.  
In this letter, we point out 
that the Jacobi identities for brackets containing such terms 
are not guaranteed to hold.  The quantity $\pa_x^{-1}\, \d (x-y)$ is not 
uniquely
determined in terms of Heaviside functions, and as shown in \rf{crucial}, that 
undeterminacy involves just one parameter. We show that the $sl(2)$ AKNS 
and DNLS systems admit three Hamiltonian structures, and we establish that the 
the Jacobi identities for such structures  
are only satisfied for one particular choice of that parameter. Moreover, we
also check that those three structures are compatible \ct{magri} and so the 
$sl(2)$ AKNS and DNLS systems are tri-Hamiltonian. In the case of $sl(3)$ AKNS
system we have shown that the choice of the above mentioned parameter, in order
for the Jacobi identity to hold, is not the same for all terms in the second
bracket, involving $\pa_x^{-1}\, \d (x-y)$. We have determined in fact, that 
there
exists two possible solutions. 
We also give a derivation of the recursion operator for the $sl(n+1)$ DNLS
system.

\sect{AKNS}

The Generalized $sl(n+1)$ AKNS theory is defined by the linear matrix problem 
 \ct{FK83}
\br
A \Psi \eq \pa \Psi \lab{1.1} \\
B_m \Psi \eq \pa_{t_m} \Psi \qquad m=2,3,\ldots \lab{1.2}
\er
for
\be
A = \l E + A^{0} \qquad {\rm with} \qquad
E = { 2 \l_n \cdot H \o \a_n^2}
\lab{1.3}
\ee
where $\l_n$ and $\a_n$ are the n-th fundamental weight and simple root of 
$sl(n+1)$ respectively, 
and\footnote{Throughout this paper we will use the Chevalley basis for a Lie 
algebra $\cg$,
$H_a$, $E_{\a}$, satisfying $\sbr{H_a}{H_b} = 0$, $\sbr{H_a}{E_{\a}} =  
K_{\a a} \,E_{\a}$ and $\sbr{E_{\a}}{E_{\b}}$ is equal to 
$\eps (\a ,\b ) E_{\a +\b}$ if $\a +\b$ is a root, to
$\sum _{a=1}^{n} n_a H_a$ if $\a +\b =0$, and $0$ otherwise. We have denoted  
$K_{\a a} = {2\a . \a_a \o{\a_a^2}} = \sum n_bK_{ba}$, with $K_{ab}$ being 
the Cartan matrix, and a root $\a$ can be expanded in terms of simple roots
as $\a = \sum n_a \a_a$. The quantities $\eps (\a ,\b )$ are integers (just
signs for simple laced algebras) determined by use of the Jacobi identities.}
\be
A^{0}= \sum_{a=1}^n \( q_a E_{(\a_a+ \ldots+\a_n)}+ r_a E_{-(\a_a+ \ldots
+\a_n )}  \)
\lab{1.11}
\ee
and where $q_a$ and $r_a$ are the fields of the model, satisfying the
Generalized Non-Linear Schrodinger equations
\br
 {\pa q_a \o{\pa t}}& =& \pa_x^2 q_a - 2q_a\sum_{b=1}^{n} \, q_b \, r_b
\nonumber \\
{\pa r_a \o{\pa t}} &=& -\pa_x^2 r_a + 2r_a\sum_{b=1}^{n} \,q_b \, r_b
\lab{nls}
\er

The model described by the eqs. \rf{nls} is a representative of a hierarchy
which consists of an infinite set of equations involving an infinite number of
times. Successive flows are related by the recursion operator $\cR$, as
explained in \ct{agz}, by   
\br
&&\pa_{t_n} \twocol{r_i}{q_l} = \cR_{(i,l),(j,m)} \pa_{t_{n-1}} 
\twocol{r_j}{q_m} =
\lab{1.19}\\
&&\fourmat{\(-\pa + r_k \pa^{-1} q_k\)\d_{ij}+r_i\pa^{-1}q_j}
{r_i \pa^{-1} r_m +r_m\pa^{-1}r_i}{-q_l \pa^{-1} q_j -q_j\pa^{-1}q_l}
{\(\pa -q_k\pa^{-1}r_k\)\d_{lm} -q_l \pa^{-1} r_m}
\pa_{t_{n-1}} \twocol{r_j}{q_m}
\nonu
\er

A first Hamiltonian structure is introduced from a Poisson bracket 
\be
P_1(x,y) =
\fourmat{\pbr{r_i}{r_j}}{\pbr{r_i}{q_m}}{\pbr{q_l}{r_j}}{\pbr{q_l}{q_m}}
= \fourmat{0}{-I}{I}{0} \d (x-y)
\lab{1.22}
\ee

A second Poisson bracket structure can be obtained from the recursion operator 
\be
\cR = P_2 P_1^{-1}
\lab{1.23}
\ee
leading to 
\br
&&P_2(x,y) = \fourmat{0}{\d_{im}}{\d_{lj}}{0} \pa_x \, \d (x-y) + 
\\
&& \fourmat{r_i(x)  r_j(y)  + r_j(x)  r_i(y)}
{ -\d_{im}\sum_k r_k(x)  q_k(y)  - r_i(x)  q_m(y)}
{ -\d_{lj}\sum_k q_k(x)  r_k(y)  - q_l(x)  r_j(y)}
{q_l(x)  q_m(y) + q_m(x) q_l(y)} \, \pa^{-1}_x \, \d (x-y) \nonu
\lab{P2}
\er

We say two brackets $\pbrs{\cdot}{\cdot}{i}$ and $\pbrs{\cdot}{\cdot}{j}$ are 
compatible if
\be
K_{ij}(A,B,C)\equiv J_{ij}(A,B,C) + J_{ji}(A,B,C) = 0
\lab{compatible}
\ee
where
\be
J_{ij}(A,B,C) \equiv \pbrs{\pbrs{A}{B}{i}}{C}{j} + \pbrs{\pbrs{C}{A}{i}}{B}{j}
+ \pbrs{\pbrs{B}{C}{i}}{A}{j} 
\ee 

As we have mentioned in the introduction the bracket $P_2$ does not
necessarily satisfy the Jacobi identities. That can be easily seen by
considering the Jacobi identity for the three fields $r_1(x)$, $r_1(y)$ and
$r_2(z)$. The source of the problem lies in the way one expresses the quantity 
$\pa^{-1}_x\d (x-y)$ in terms of Heaviside's $\Theta$ functions. The generic
form of such relation is 
\be
\pa^{-1}_x\d (x-y) = \g \Th (x-y) - (1 -\g ) \Th (y-x)
\lab{crucial}
\ee
since (see \rf{thetader})
\be
\pa_x \, \pa^{-1}_x\d (x-y) = \d (x-y) \qquad \mbox{\rm for any $\g$}
\ee

\subsection{The $sl(2)$ case}

As a first example consider the $sl(2)$ case, where the first and second
brackets are given respectively by
\br
\pbrs{r(x)}{r(y)}{1} = \pbrs{q(x)}{q(y)}{1} = 0 \qquad 
\pbrs{r(x)}{q(y)}{1}= - \d (x-y)
\er
and (from now on we will use primes to denote derivatives w.r.t. the argument
of the function, e.g. $\pa_x \, \d (x-y) \equiv \d^{\pr} (x-y)$)
\br
\pbrs{r(x)}{r(y)}{2} &=& 2 r(x) \, r(y) \, \pa^{-1}_x \d (x-y)  
\lab{sl2bra2a}\\
\pbrs{q(x)}{q(y)}{2} &=& 2 q(x) \, q(y) \, \pa^{-1}_x \d (x-y) 
\lab{sl2bra2b} \\
\pbrs{r(x)}{q(y)}{2} &=&  \d^{\pr} (x-y) - 2 r(x) \, q(y) \, \pa^{-1}_x 
\d (x-y)
\lab{sl2bra2c}
\er

Since the bracket has to be antisymmetric under the exchange of its entries, we
choose, on \rf{sl2bra2a} and \rf{sl2bra2b}, the constant $\g$ introduced in
\rf{crucial} to be $\h$. Then the Jacobi identity implies that we have to make
the same value for $\g$ in \rf{sl2bra2c}. 

We can also introduce a third bracket by the relation
\be
P_3 \equiv P_2 \, P_1^{-1} P_2
\lab{p2p1p2}
\ee
giving
\br
\pbrs{r(x)}{r(y)}{3} &=& - \( r(x) \, r^{\pr}(y) + r^{\pr}(x)\, r(y) \) 
\eps (x-y)\\
\pbrs{q(x)}{q(y)}{3} &=&  \( q(x) \, q^{\pr}(y) + q^{\pr}(x)\, q(y) \) 
\eps (x-y)\\
\pbrs{r(x)}{q(y)}{3} &=& -\d^{\pr\pr}(x-y) - 
\( r(x) \, q^{\pr}(y) - r^{\pr}(x)\, q(y) \) 
\eps (x-y) \nonu\\
&+& 4 r(x) \, q(x) \, \d (x-y)
\er

We have checked, using the Mathematica program, that the three brackets defined 
above
satisfy the Jacobi identities, and in addition, that they are all compatible
with each other, in the sense of \rf{compatible}. 
The above results establish that the $sl(2)$-AKNS
system is tri-Hamiltonian. 

\subsection{The $sl(3)$ case}

We have checked that in order for the Jacobi identity to be satisfied, the 
second bracket for the $sl(3)$-AKNS system given in \rf{P2}, one can not choose
the  value of $\g$ introduced in \rf{crucial}, to be the same for all brackets.
Because of antysymmetry, one has to take $\g =\h$ for brackets involving the 
same type of fields, and for the other brackets we determine the values of $\g$
imposing the Jacobi identity. The result we got, using Mathematica, is that
there are two possibilities for the choices of $\g$,  specified below
by the parameter $\eta$ which can take the values $\pm 1$.
\br
\pbrs{r_i(x)}{r_i(y)}{2} &=& r_i(x)\, r_i(y)\, \eps (x-y) \qquad 
\mbox{\rm for $i=1,2$}\\
\pbrs{q_i(x)}{q_i(y)}{2} &=& q_i(x)\, q_i(y)\, \eps (x-y) \qquad 
\mbox{\rm for $i=1,2$}\\
\pbrs{r_i(x)}{q_i(y)}{2} &=& \d^{\pr}(x-y) - r_i(x)\, q_i(y) \, \eps (x-y) 
\nonu\\
&+& \eta \, \sum_{j=1}^{2}\varepsilon_{ij} \, r_j(x)\, q_j(y) \, 
\Th (\eta \varepsilon_{ij}(y-x))\qquad 
\mbox{\rm for $i=1,2$}\\
\pbrs{r_1(x)}{r_2(y)}{2} &=& r_1(x)\, r_2(y)\, \Th (x-y) - 
\eta \, r_1(y)\, r_2(x) \, \Th (\eta (y-x))\\
\pbrs{q_1(x)}{q_2(y)}{2} &=& q_1(x)\, q_2(y)\, \Th (x-y) + 
\eta \, q_1(y)\, q_2(x) \, \Th (\eta (x-y))\\
\pbrs{r_1(x)}{q_2(y)}{2} &=& - r_1(x)\, q_2(y)\, \Th (x-y)\\
\pbrs{r_2(x)}{q_1(y)}{2} &=&  r_2(x)\, q_1(y)\, \Th (y-x)
\er
where $\varepsilon_{ij}$ is antisymmetric and $\varepsilon_{12} =1$.

It is now straightforward to show that the first bracket defined in \rf{1.22}
is compatible with the above second brackets for the two choices of $\eta$,
namely $\eta = \pm 1$. Therefore the $sl(3)$-AKNS system is bi-Hamiltonian.  
Recently, Liu \ct{liu} has argued that the $sl(3)$-AKNS system is 
bi-Hamiltonian. However, it was not taken into account the fact that the Jacobi
identity may not be satisfied, if a careful prescription for $\pa^{-1}_x \d
(x-y)$ (see \rf{crucial}) is not made. 

Let us mention that through \rf{p2p1p2} we can calculate the third bracket for
this case. However, we need to check the Jacobi identity and the compatibility
condition \rf{compatible} with the other two brackets in order to establish
that the system is indeed tri-Hamiltonian. 

\sect{DNLS}

The Generalized $sl(n+1)$-DNLS system is defined by \rf{1.1}-\rf{1.2} for $A$
given by \ct{fordy}
\be
A = \l^2 E + \l A^{0}
\ee
with $E$ given by \rf{1.3}, and 
\be
A^{0}= \sum_{a=1}^n \( - q_a E_{(\a_a+ \ldots+\a_n)}+ r_a E_{-(\a_a+ \ldots
+\a_n )}  \)
\ee
The corresponding Generalized DNLS equations of motion are
\br
 {\pa q_a \o{\pa t}}& =& \pa_x^2 q_a + 
 2\pa_x\(q_a\sum_{b=1}^{n} \, q_b \, r_b\)
\nonumber \\
{\pa r_a \o{\pa t}} &=& -\pa_x^2 r_a + 
2\pa_x\( r_a\sum_{b=1}^{n} \,q_b \, r_b\)
\lab{dnls}
\er

Let us now discuss the recursion operator for such system. Consider the
Zakharov-Shabat equations for consecutive times $t_m$ and $t_{m-1}$
\br
\pa_{t_m} \, A - \pa_x\, B_m + \sbr{A}{B_m} &=& 0 
\lab{zs1}\\ 
\pa_{t_{m-1}} \, A - \pa_x\, B_{m-1} + \sbr{A}{B_{m-1}} &=& 0
\lab{zs2}
\er
In order to determine the general solution of these equations we take the
ansatz
\be
B_m = \l^2 \, B_{m-1} + \l^2 \, C_m + \l \, D_m + Y_m
\lab{ansatz}
\ee
Multiplying \rf{zs1} by $\l$, \rf{zs2} by $\l^3$, taking the difference and
using the ansatz \rf{ansatz},  one gets
\br
\l \, \pa_{t_m} \, A^{0} &-& \l^3 \, \pa_{t_{m-1}} \, A^{0} - 
\pa_x \(\l^2 \, C_m + \l \, D_m + Y_m\) - 
\l^2 \,\sbr{E}{\l^2 \, C_m + \l \, D_m +Y_m} \nonu\\
&+& 
\l\, \sbr{A^{0}}{\l^2 \, C_m + \l \, D_m + Y_m} = 0
\er
The $\l$ independent components yields 
\br
\pa_x\, Y_m =0
\er
so that we can choose $Y_m =0$. 
The other components yields the following equations
\br
\pa_{t_m} \, A^{0}  - \pa_x \, D_m &=& 0 
\lab{eqa}\\
\pa_x \, C_m - \sbr{A^{0}}{D_m} &=& 0 
\lab{eqb}\\
\pa_{t_{m-1}} \, A^{0} - \sbr{E}{D_m} - \sbr{A^{0}}{C_m} &=& 0 
\lab{eqc}\\
\sbr{E}{C_m} &=& 0 
\lab{eqd}
\er
{}From the last equation we conclude that $C_m \in\kere$. Since $A^{0}\in
\ime$, the first equation implies that also $D_m \in \ime$ (except for a 
$x$ independent component in $\kere$ which we do not consider). 

\subsection{The $sl(2)$ case}

In this case one has\footnote{Here $\sigma_{3,\pm}$ stands for the combinations
of the Pauli
matrices such that, $\sbr{\sigma_3}{\sigma_{\pm}}= \pm 2 \sigma_{\pm}$ and 
$\sbr{\sigma_+}{\sigma_-}= \sigma_3$.}
\be
E = \h\, \sigma_3 \, , \qquad A^{0} = - q \, \sigma_+ + r\, \sigma_-
\lab{esl2}
\ee
and so, since $C_m \in \kere$ and $D_m \in \ime$ one has
\be
C_m = c_m \, \sigma_3 \, , \qquad D_m =  d_m^+\, \sigma_+ + d_m^- \, \sigma_- 
\ee

Replacing \rf{esl2} into \rf{eqb} and \rf{eqc} one gets
\br
\pa_x \, c_m + q \,d_m^- + r \,  d_m^+ &=& 0\lab{eq2a}\\
\pa_{t_{m-1}} \, q - d_m^+ - 2 q \, c_m &=& 0\lab{eq2b}\\ 
\pa_{t_{m-1}} \, r + d_m^- \, - 2 r c_m &=& 0 \lab{eq2c}
\er
{}From these equations one finds 
\be
c_m = \pa_x^{-1} \, \pa_{t_{m-1}} \( r\, q\)
\ee
Replacing into \rf{eq2b} and \rf{eq2c} leads to 
\br
d_m^+  &=& - \pa_{t_{m-1}} \, q - 
2 q \, \pa_x^{-1} \, \pa_{t_{m-1}} \( r\, q\) \\
d_m^-  &=& - \pa_{t_{m-1}} \, r + 
2 r \, \pa_x^{-1} \, \pa_{t_{m-1}} \( r\, q\) 
\er
Substituting into \rf{eqa}, one can write it in the form  of \rf{1.19} with the
recursion operator given by
\br
R(x,y) &=& \fourmat{-1}{0}{0}{1} \, \pa_x \, \d (x-y) + 
2\,\fourmat{ q(x) r(x)}{r(x)^2}{q(x)^2}{ q(x) r(x)}\, \d (x-y) \nonu\\
&+&
2\, \fourmat{r^{\pr}(x)q(y)}{r^{\pr}(x)r(y)}{q^{\pr}(x)q(y)}{q^{\pr}(x)r(y)} \,
\pa_x^{-1}\, \d (x-y)
\er
One can verifies that the recursion operator can be decomposed into
\be
R = P_3\, P_2^{-1}
\ee
where 
\br
P_3(x,y)= \(
\begin{array}{rr}
0 & \d^{\pr}(x-y)\\
\d^{\pr}(x-y) & 0
\end{array}\)
\er
and
\br
P_2(x,y)= \fourmat{0}{-1}{1}{0}\, \d (x-y) + 
2\, \fourmat{r(x)r(y)}{-r(x)q(y)}{-q(x)r(y)}{q(x)q(y)}\, \pa_x^{-1}\, \d (x-y)
\lab{p2pr}
\er
The brackets associated to the above $P_2$ and $P_3$ were proposed in
\ct{swieca}, by choosing the prescription $\g = \h$ in \rf{crucial}, i.e. 
\be
2\, \pa_x^{-1}\, \d (x-y) =  \eps (x-y)
\ee
in all entries of \rf{p2pr}.

We can obtain a further bracket, denoted $P_4$, by 
\be
R = P_4 \, P_3^{-1}
\lab{p3p4}
\ee
It gives
\br
\pbrs{r(x)}{r(y)}{4} &=& - r^{\pr}(x)\, r^{\pr}(y) \, \eps (x-y) + 
2 r^{\pr}(x)\, r(y) \, \d (x-y) + 2 r(x)^2 \, \d^{\pr}(x-y)
\lab{bra4a}\\
\pbrs{q(x)}{q(y)}{4} &=& - q^{\pr}(x)\, q^{\pr}(y) \, \eps (x-y) + 
2 q^{\pr}(x)\, q(y) \, \d (x-y) + 2 q(x)^2 \, \d^{\pr}(x-y)
\lab{bra4b}\\
\pbrs{r(x)}{q(y)}{4} &=& - \d^{\pr\pr}(x-y) - 
r^{\pr}(x)\, q^{\pr}(y) \, \eps (x-y) + 
2 r^{\pr}(x)\, q(x) \, \d (x-y) \nonu\\
&+&
2 r(x)\, q(x) \, \d^{\pr}(x-y)
\lab{bra4c}
\er

The third and fourth brackets, $P_3$ and $P_4$, were introduced in ref.
\ct{willox}, from which they constructed the recursion operator $R$ through 
\rf{p3p4}. 

We have verified, using the Mathematica program, that the three brackets 
$P_2$, $P_3$ and $P_4$ satify the Jacobi
identities and are compatible with each other in the sense of \rf{compatible}. 
Therefore, we have established that such system is tri-Hamiltonian. 

\subsection{The $sl(3)$ case}

Let us parametrize, in this case, the quantities introduced in \rf{ansatz} as
\br
C_m &=& c_m^{(\a_1)} \, E_{\a_1} + c_m^{(-\a_1)} \, E_{-\a_1} + 
c_m^{(1)} \, H_1 + c_m^{(2)} \, H_2 \\
D_m &=& d_m^{(\a_1+\a_2)} \, E_{\a_1+\a_2} + d_m^{(\a_2)} \, E_{\a_2} + 
d_m^{(-\a_1-\a_2)} \, E_{-\a_1-\a_2} + d_m^{(-\a_2)} \, E_{-\a_2}\\
A^{0} &=& - q_1 \, E_{\a_1+\a_2} - q_2 \, E_{\a_2} + r_1 \, E_{-\a_1-\a_2} + 
r_2 \, E_{-\a_2}
\er
Substituting in \rf{eqb} one gets
\br
\pa_x \, c_m^{(\a_1)} &=& -q_1 \, d_m^{(-\a_2)} - r_2 \,d_m^{(\a_1+\a_2)}
\lab{difeq1}\\
\pa_x \, c_m^{(-\a_1)} &=& - q_2 \, d_m^{(-\a_1-\a_2)} - r_1 \, d_m^{(\a_2)}
\lab{difeq2}\\
\pa_x \, c_m^{(1)} &=&  -q_1 \, d_m^{(-\a_1-\a_2)} - r_1 \,d_m^{(\a_1+\a_2)}
\lab{difeq3}\\
\pa_x \, c_m^{(2)} &=&  -q_1 \, d_m^{(-\a_1-\a_2)} - r_1 \,d_m^{(\a_1+\a_2)} 
- q_2 \, d_m^{(-\a_2)} - r_2 \, d_m^{(\a_2)}
\lab{difeq4}
\er
whilst \rf{eqc} leads to
\br
 d_m^{(\a_1+\a_2)}&=&-\pa_{t_{m-1}} \, q_1  - q_1 \,\( c_m^{(1)} + c_m^{(2)}\) 
- q_2 \,c_m^{(\a_1)} 
\lab{difeq5}\\ 
d_m^{(\a_2)}&=&-\pa_{t_{m-1}} \, q_2   - q_2 \,\( - c_m^{(1)} + 2 c_m^{(2)}\) - 
q_1 \,c_m^{(-\a_1)}  
\lab{difeq6}\\
d_m^{(-\a_1-\a_2)}&=&-\pa_{t_{m-1}} \, r_1   + r_1 \, \( c_m^{(1)} + c_m^{(2)}\) 
 + r_2 \,c_m^{(-\a_1)} 
 \lab{difeq7}\\
d_m^{(-\a_2)}&=&-\pa_{t_{m-1}} \, r_2   + r_2 \,\( - c_m^{(1)} + 2 c_m^{(2)}\) 
+ r_1 \,c_m^{(\a_1)}  
\lab{difeq8}
\er
{}From the eqs. \rf{difeq1}-\rf{difeq8} one gets
\be
c_m^{(2)} = \pa_x^{-1}\, \pa_{t_{m-1}} \( r_1 q_1 + r_2 q_2 \)
\ee
and
\br
\pa_x \, c_m^{(\a_1)} -  q_1 r_2 \( 2 c_m^{(1)} - c_m^{(2)} \)  -  
\( r_2q_2 - r_1q_1\) c_m^{(\a_1)}  &=& \pa_{t_{m-1}} \,\( q_1 r_2\)
\lab{sys1}\\
\pa_x \, c_m^{(-\a_1)} +  q_2 r_1 \( 2 c_m^{(1)} - c_m^{(2)} \)  +  
\( r_2q_2 - r_1q_1\) c_m^{(-\a_1)}  &=& \pa_{t_{m-1}} \,\( q_2 r_1\)
\lab{sys2}\\
\pa_x \, c_m^{(1)} - r_1 q_2 c_m^{(\a_1)} + r_2 q_1 c_m^{(-\a_1)} &=& 
\pa_{t_{m-1}} \,\( q_1 r_1\) 
\lab{sys3}
\er

One can write the above system of differential equations in a more elegant way,
introducing 
\br
c_m^{\pm} \equiv c_m^{\a_1} \pm c_m^{-\a_1}\, , \qquad 
c_m^{0} \equiv 2 c_m^{1} - c_m^{2}
\er
and also
\br
\g_{\pm} \equiv q_1\, r_2 \pm q_2 \, r_1 \, , \qquad 
\g_0 \equiv q_1\, r1 - q_2 \, r_2
\er
Now, introduce the matrices
\br
c\equiv \( \begin{array}{c}
c_+ \\
c_0\\
c_-
\end{array}\) \, ; \qquad
\g \equiv \( \begin{array}{c}
\g_+ \\
\g_0\\
\g_-
\end{array}\) 
\er
and
\br
W\equiv \( \begin{array}{ccc}
0 & - \g_- & \g_0\\
\g_- & 0 & - \g_+\\
\g_0 & - \g_+ & 0
\end{array}\) 
\er

Then, the solution for the system \rf{sys1}-\rf{sys3} can be written 
formally as 
\be
c = D_x^{-1} \, \pa_{t_{m-1}} \, \g \, \qquad D_x \equiv \pa_x + W
\ee
Notice that $W$ is a matrix in the adjoint of $sl(2)$, and therefore in the 
actual integration of the above equations such algebraic structure should play
an important role.

\sect{Appendix}

We give here some definitions and relations involving delta and 
Heaviside functions.  

The Heaviside function is defined by
\br
\Th (x-y) \equiv \lcurl  
\begin{array}{ll}
1 & \mbox{\rm for $x>y$}\\
0 & \mbox{\rm for $x<y$}\\
\h & \mbox{\rm for $x=y$}
\end{array} \right. 
\lab{thetadef}
\er

We also introduce the sign function as
\be
\epsilon (x-y) \equiv \Th (x-y) - \Th (y-x)
\lab{epsdef}
\ee 
One also has 
\be
\pa_x \, \Th (x-y) = - \pa_y \, \Th (x-y) = \d (x-y)
\lab{thetader}
\ee
and so
\be
\pa_x \, \epsilon (x-y) = - \pa_y \, \epsilon (x-y) = 2 \d (x-y)
\lab{epsder}
\ee

In order to verifiy Jacobi identities and other relations involving fields at
different points we used the strategy of using delta functions and its
derivatives to try to write the fields at the same point. The relations used 
can be derived from the identity
\be
f(y) \, \d (x-y) = f(x) \, \d (x-y)
\ee
Indeed, differentiating it one gets 
\be
f(y) \, {d^n \,\d (x-y) \o d\, x^n} = 
\sum_{l=0}^n \twocol{n}{l}\, {d^{n-l} \,\d (x-y) \o d\, x^{n-l}} \, 
{d^l \, f(x) \o d\, x^l}
\lab{useful}
\ee
We also have used the identities
\br
\Theta (y-x)\, \Theta (z-x) - \Theta (y-z)\, \Theta (z-x) - 
\Theta (y-x)\, \Theta (z-y) &=& 0 \nonu\\
\Theta (x-y)\, \Theta (z-x) + \Theta (y-x)\, \Theta (z-y) - 
\Theta (z-x)\, \Theta (z-y) &=& 0 
\er

\vspace{.5 cm}

\noindent{\bf Acknowledgements}

We are greatful to H. Aratyn for the correspondence on the subject.

\end{document}